\documentclass[a4paper]{jpconf}
\usepackage{graphicx}
\def\lsim{\raise0.3ex\hbox{$<$\kern-0.75em\raise-1.1ex\hbox{$\sim$}}}
\def\gsim{\raise0.3ex\hbox{$>$\kern-0.75em\raise-1.1ex\hbox{$\sim$}}}
\begin{document}
\title{Lattice QCD results on cumulant ratios at freeze-out}

\author{Frithjof Karsch}

\address{
Fakult\"at f\"ur Physik, Universit\"at Bielefeld, D-33615 Bielefeld,
Germany;\\
Physics Department, Brookhaven National Laboratory, Upton, NY 11973, USA}

\ead{karsch@bnl.gov}

\begin{abstract}
Ratios of cumulants of net proton-number fluctuations
measured by the STAR Collaboration show strong deviations from a skellam
distribution, which should describe thermal properties of cumulant ratios,
if proton-number fluctuations are generated in
equilibrium and a hadron resonance gas (HRG) model would provide a suitable
description of thermodynamics at the freeze-out temperature.
We present some results on $6^{th}$ order cumulants 
entering the calculation of the QCD equation of state at non-zero values of the 
baryon chemical potential ($\mu_B$) and
discuss limitations on the applicability of HRG thermodynamics deduced from
a comparison between QCD and HRG model calculations of cumulants 
of conserved charge fluctuations. We show that basic features of
the $\mu_B$-dependence of skewness and kurtosis ratios of net proton-number fluctuations measured by the STAR Collaboration
resemble those expected from a ${\cal O}(\mu_B^2)$ QCD calculation 
of the corresponding net baryon-number cumulant ratios.

\end{abstract}

\section{Introduction}
A major goal in current experimental and theoretical studies of the
thermodynamics of strong interaction matter is the exploration of its
phase diagram. The hope is to find evidence for the existence of a
second order phase transition point -- the chiral critical point (CCP) -- 
located at some value of the chemical potential, $\mu_B^{crit}$ 
\cite{Ding:2015ona}. This 
would be the starting point for a line of first order phase transitions
at larger values of the baryon chemical potential $\mu_B$. 

At RHIC a dedicated research program -- the beam energy scan (BES) -- 
has been established that seeks evidence for the existence and location 
of the CCP. By varying the beam energy
properties of matter in a regime of temperatures ($T$) up to about three
times the transition temperature, $T_{pc}\sim 155$~MeV \cite{pap3,Bazavov:2011nk},
and baryon chemical potential up to $\mu_B\simeq 3T$ can be probed.
It is generally expected that conserved charge fluctuations, which are
generated close to, or at the freeze-out temperature, $T_f(\mu_B)$, can provide
insight into the existence and location of the CCP. An important prerequisite
for such studies, however, is to understand the thermodynamics of 
hot and dense matter in the crossover region and, in particular, close
to freeze-out in QCD. 

In the following we will point out the importance
of characterizing this regime in terms of QCD rather than hadron resonance
gas (HRG) model calculations, which are quite successful in approximating
QCD thermodynamics at sufficiently low temperatures, but definitely fail
to capture important aspects of QCD thermodynamics visible in conserved
charge fluctuations at temperatures $T\gsim 160$~MeV. 

\section{Energy density in the crossover region}
At small values of $\mu_B$ the QCD transition is not a true phase 
transition but a smooth transition from the low-$T$ hadronic
to the high-$T$ partonic regime. This {\it crossover transition} 
does not happen at a well defined temperature. However, it can be characterized
by pseudo-critical temperatures, i.e. temperatures that reflect 
characteristic
features of e.g. fluctuation observables, which are guaranteed to converge
to the true critical temperature in the chiral limit.
One such observable is the chiral susceptibility, $\chi_q$,
the derivative of the chiral condensate with respect to quark mass. The location of the maximum
of $\chi_q$
defines a pseudo-critical
temperature. This has been obtained in lattice QCD calculations, 
$T_{pc} = 154(9)$~MeV \cite{Bazavov:2011nk}. 

\begin{figure}[t]
\includegraphics[width=7.5cm]{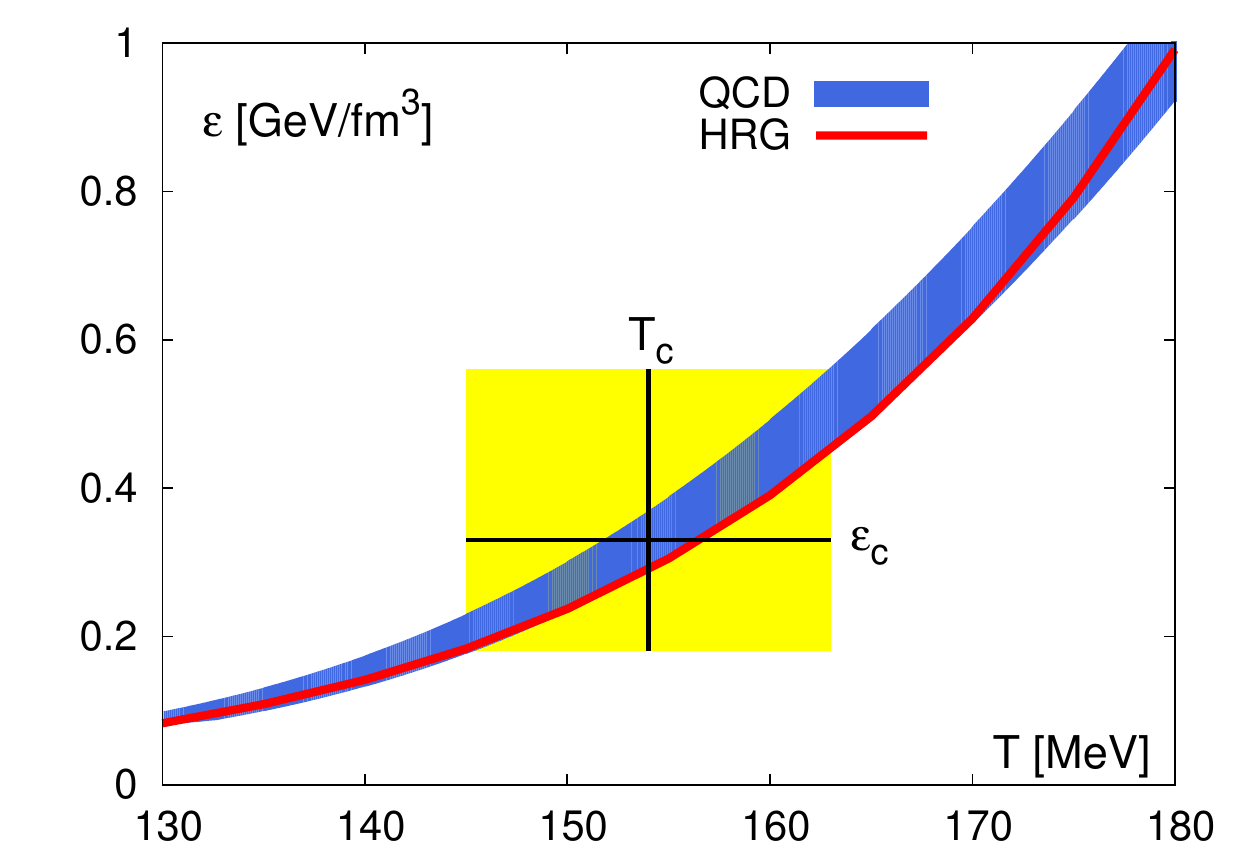}
\includegraphics[width=7.7cm,height=5cm]{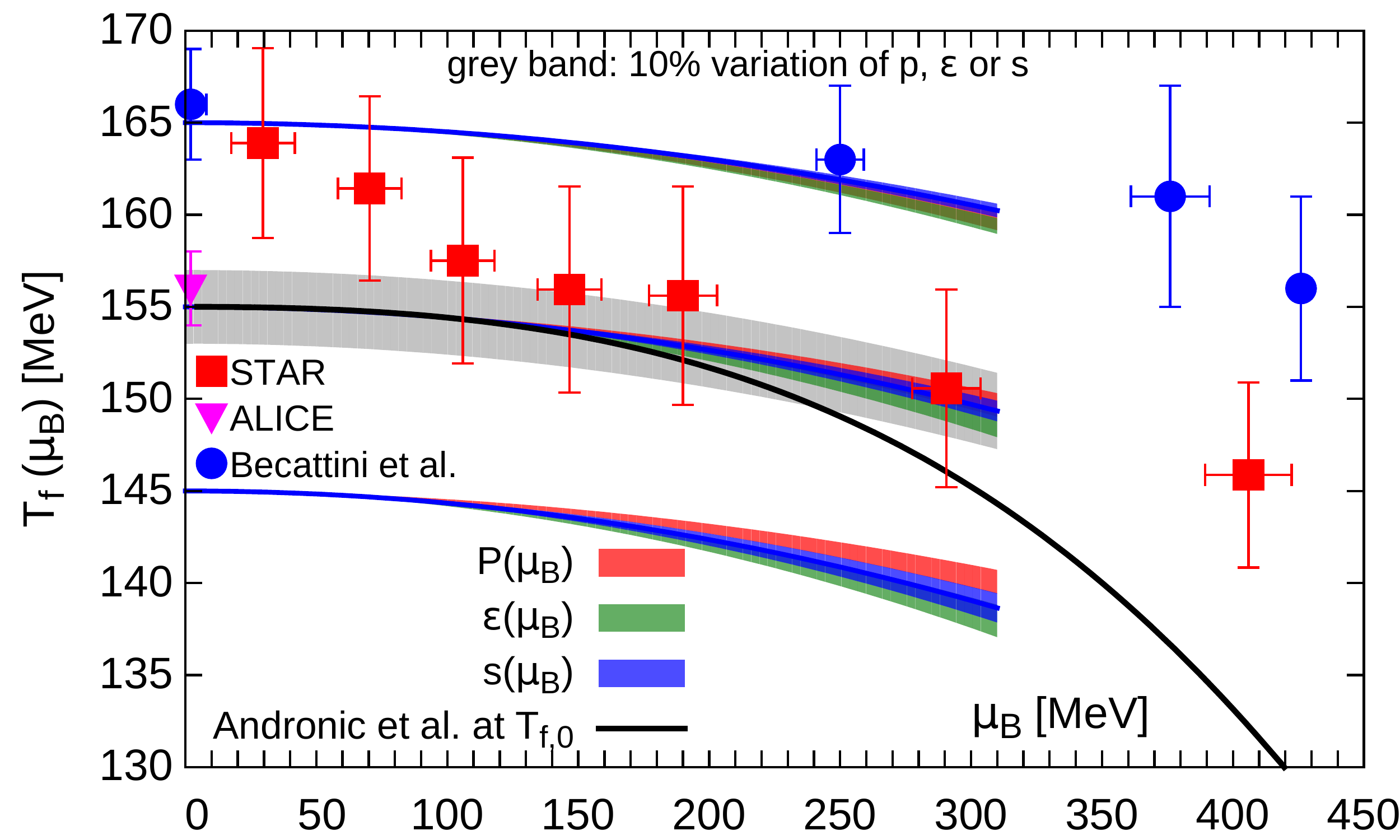}
\caption{\label{fig:energy}
{\it Left}: Energy density of QCD versus temperature obtained
from lattice QCD calculations (band) and HRG model
calculations (line).
{\it Right}: Freeze-out temperatures versus baryon chemical
potential determined by STAR (boxes) \cite{Das:2014qca} and ALICE (triangle) 
\cite{Floris:2014pta}, and 
hadronization temperatures (circle) determined in Ref.~\cite{Becattini:2016xct}.
The bands show lines of constant pressure, energy density and entropy density,
respectively. They have been determined from a ${\cal O}(\mu_B^2)$ Taylor 
expansion of the QCD partition function. 
Also shown is a parametrization of the freeze-out line
suggested by Andronic et al. (black line) \cite{Andronic:2008gu} shifted 
to $T_{f}(0)=155$~MeV.}
\end{figure}

In Fig.~\ref{fig:energy}~(left) we show results for the energy density 
($\epsilon$) obtained from lattice QCD calculations with physical strange and 
light quark masses \cite{Bazavov:2014pvz}. The temperature range covered by 
current uncertainties in $T_{pc}$ corresponds to  
$\epsilon_c= (0.34\pm 0.16)$~GeV/fm$^3$. Unfortunately, this energy density 
range is still quite large. In fact, it covers practically all
energy scales of interest. While the central value corresponds to the 
close packing limit of nucleons with a radius 
$r_n\simeq 0.8$~fm, the lower limit is close to the energy density of nuclear 
matter 
($\epsilon_{nm} \simeq 0.15$~GeV/fm$^3$) and the upper limit is larger than 
the energy density inside a nucleon ($\epsilon_{n} \simeq 0.45$~GeV/fm$^3$).

Using the Taylor expansion of the QCD equation of state one can follow
lines of constant physics in the $T$-$\mu_B$ plane. In 
Fig.~\ref{fig:energy} (right) we show lines of constant pressure ($p$),
energy density ($\epsilon$) and entropy density ($s$) 
obtained in order $\mu_B^2$. Corrections to 
this are small for $\mu_B/T\le 2$. The three sets of curves, characterizing
the current uncertainty band on $T_{pc}$, correspond 
to $\epsilon = 0.2$, $0.35$ and $0.56$~GeV/fm$^3$.
Given the current uncertainties on $T_{pc}$ 
it is not too surprising that all results on freeze-out parameters 
$(T_f(\mu_B),\mu_B)$ obtained from the analysis of particle yields at the 
LHC \cite{Floris:2014pta} and RHIC \cite{Das:2014qca} and even the rather 
large hadronization temperatures extracted 
in Ref.~\cite{Becattini:2016xct} allow to state that ''hadronization and
freeze-out of hadrons occur close to, or in the QCD crossover region''
(see Fig.~\ref{fig:energy}~(right)) 
even though the temperatures in question are quite different,
e.g. ranging from 155~MeV to 165~MeV at $\mu_B=0$. They refer
to environments, in which hadronization and freeze-out may
take place, that are quite different, e.g. the energy and entropy
density may differ by a factor 2. 
Lattice QCD calculations provide a wealth
of other observables, e.g. higher order cumulants of conserved charge 
fluctuations, that are sensitive to the changes in 
physical properties of hot and dense matter that occur in this 
temperature interval. 
Understanding these changes also is of importance for our 
understanding of freeze-out conditions determined with the BES at RHIC
as well as at the LHC.

\section{Cumulants of conserved charge fluctuations}

At $\mu_B=0$ the energy density calculated in QCD as  well as HRG models
shown in Fig.~\ref{fig:energy}~(left) varies smoothly
as function of temperature when traversing the crossover region.
Even quantitatively the latter is in quite
good agreement with QCD calculations, although it has been pointed out
that the QCD energy density is systematically larger, which may be taken
as evidence for additional hadronic degrees of freedom contributing to
bulk thermodynamics close to $T_{pc}$ 
\cite{Majumder:2010ik,Bazavov:2014xya}.
Nonetheless, 
HRG model calculations seem to describe bulk thermodynamics quite well
even at temperatures as large as $T\sim 180$~MeV. In fact, the
situation is similar even for the specific heat \cite{Bazavov:2014pvz}.
Does this mean that the strongly interacting medium can 
be described in terms of hadronic degrees of freedom in the entire crossover 
region and even above?

\begin{figure}[t]
\includegraphics[width=7.5cm]{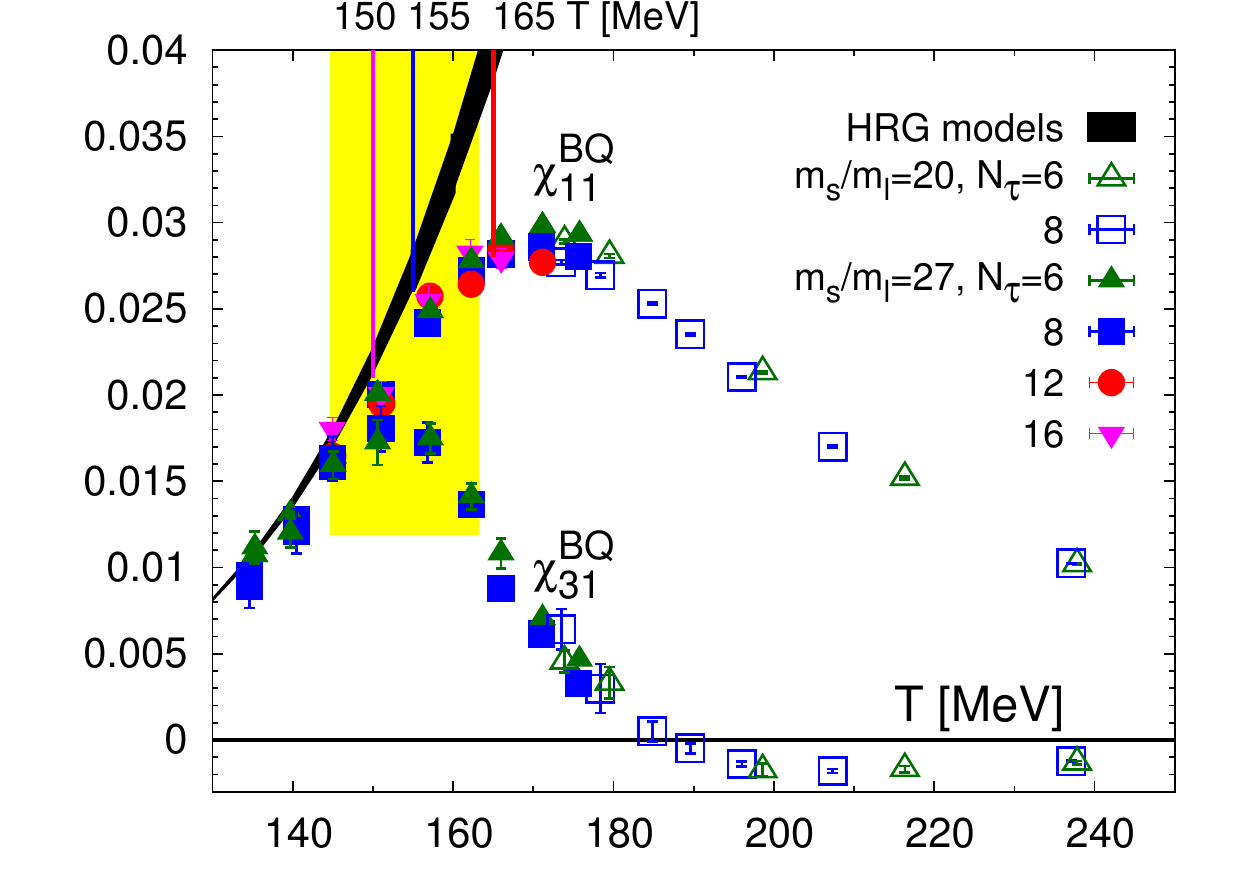}
\includegraphics[width=7.5cm]{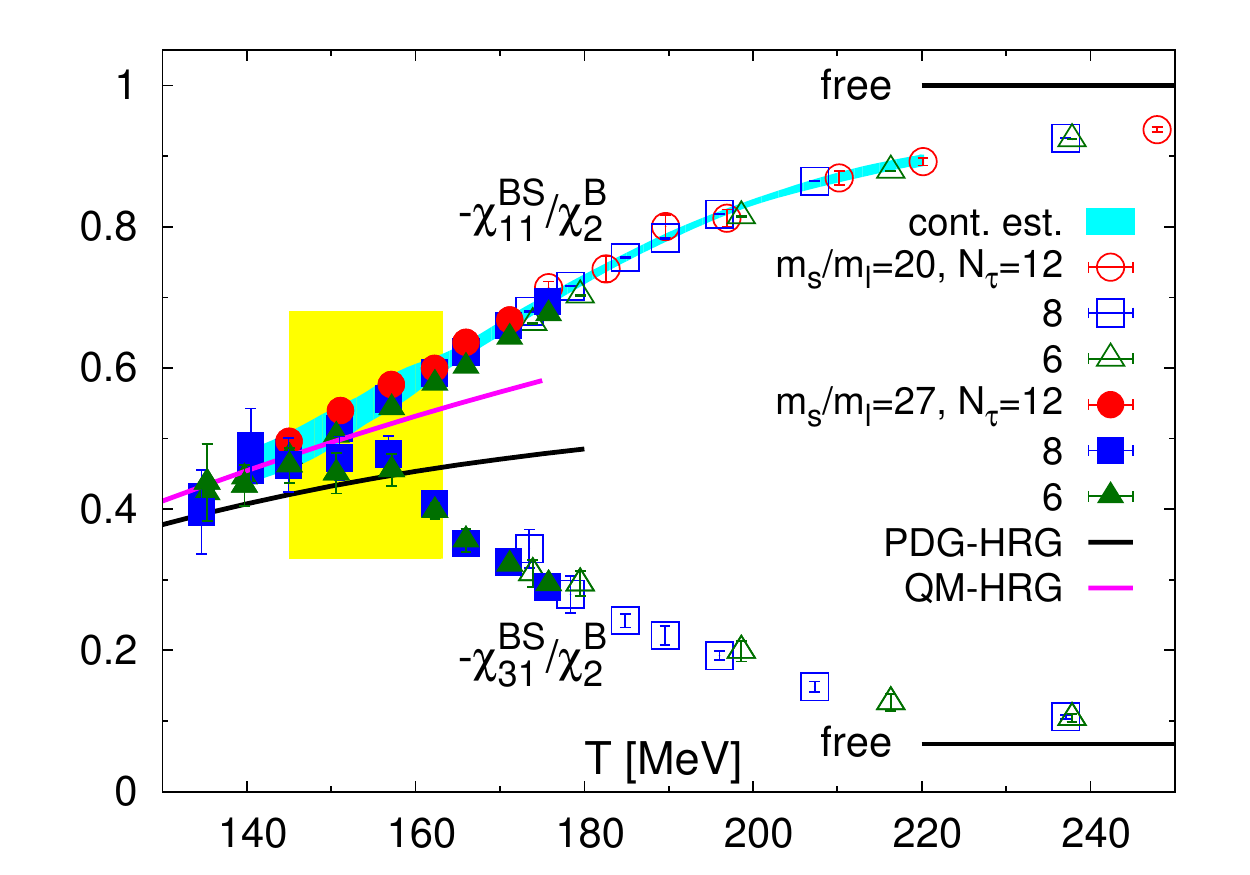}
\caption{\label{label}
{\it Left}: Correlation between net baryon-number and net electric charge
fluctuations ($\chi_{11}^{BQ}$) versus temperature
and the third moment of net baryon-number and net electric charge fluctuations
($\chi_{31}^{BQ}$). The yellow band shows the crossover region defined by
the uncertainty of the pseudo-critical temperature $T_{pc}=(154\pm 9)$~MeV.
{\it Right}: same as on the left but for correlations between net baryon-number
and net strangeness fluctuations. These cumulants have been normalized with
the quadratic fluctuations of net baryon-number.
}
\label{fig:fluctuations}
\end{figure}

Conserved charge fluctuations and correlations between them provide 
plenty of evidence that thermodynamics described in terms of 
hadronic degrees of freedom breaks down close to $T_{pc}$. In particular,
higher order cumulants are quite different from conventional HRG model
calculations. We show in Fig.~\ref{fig:fluctuations} two examples of this.
The left hand figure shows second and fourth order cumulants of 
net baryon-number and net electric charge fluctuations,
\begin{eqnarray}
\chi_{11}^{BQ} =\left.  \frac{\partial^2 p/T^4}{\partial \hat{\mu}_B \partial \hat{\mu}_Q}\right|_{\vec{\mu}=0} 
\;\; ,\;\; 
\chi_{31}^{BQ} =\left. \frac{\partial^4 p/T^4}{\partial \hat{\mu}_B^3 \partial \hat{\mu}_Q}\right|_{\vec{\mu}=0} 
\; ,
\end{eqnarray}
with $\hat{\mu}_X\equiv \mu_X/T$ and $\vec{\mu}=(\mu_B,\ \mu_Q,\ \mu_S)$. 
In the infinite
temperature, ideal quark gas limit the net electric charge of 3-flavor QCD 
vanishes. Thus $\chi_{11}^{BQ}$ and  $\chi_{31}^{BQ}$ will both approach zero
while these correlations will keep increasing exponentially in 
HRG model calculations with point-like, non interacting hadrons.

At low temperature all hadronic degrees of freedom will either carry baryon
number $B=0$ or $B=\pm 1$. HRG model calculations thus 
give  $\chi_{11}^{BQ}=\chi_{31}^{BQ}$. It is obvious from 
Fig.~\ref{fig:fluctuations}~(left)  that this relation no longer holds 
for $T\gsim 150$~MeV. Also $\chi_{11}^{BQ}$ starts deviating from HRG model
calculations at $T\gsim 155$~MeV. A similar pattern is found for second and
fourth order cumulants of net baryon-number and net strangeness correlations
shown in Fig.~\ref{fig:fluctuations}~(right). Various other combinations
of 
$2^{nd}$ and $4^{th}$ order cumulants have been constructed that make it
apparent that HRG models with point-like non-interacting hadrons are not
suitable for describing QCD thermodynamics at temperatures $T\gsim T_{pc}$
\cite{Bazavov:2013dta}. In fact, this also has been verified for
net baryon-number and net charm correlations \cite{Bazavov:2014yba} which 
strongly suggests that also the thermodynamics of open charm baryons cannot 
be described by HRG models above $T_{pc}$.

\begin{figure}[t]
\includegraphics[width=7.5cm]{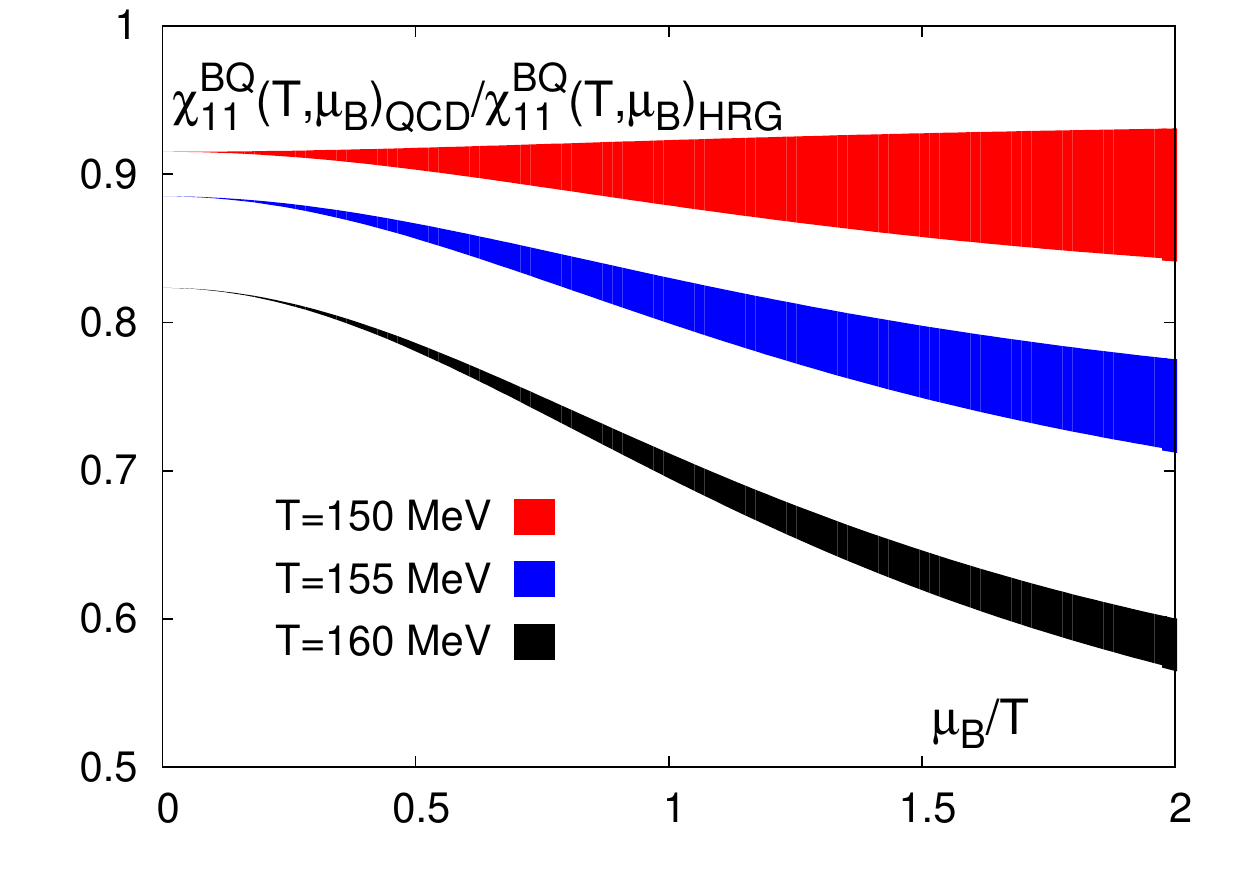}
\includegraphics[width=7.9cm]{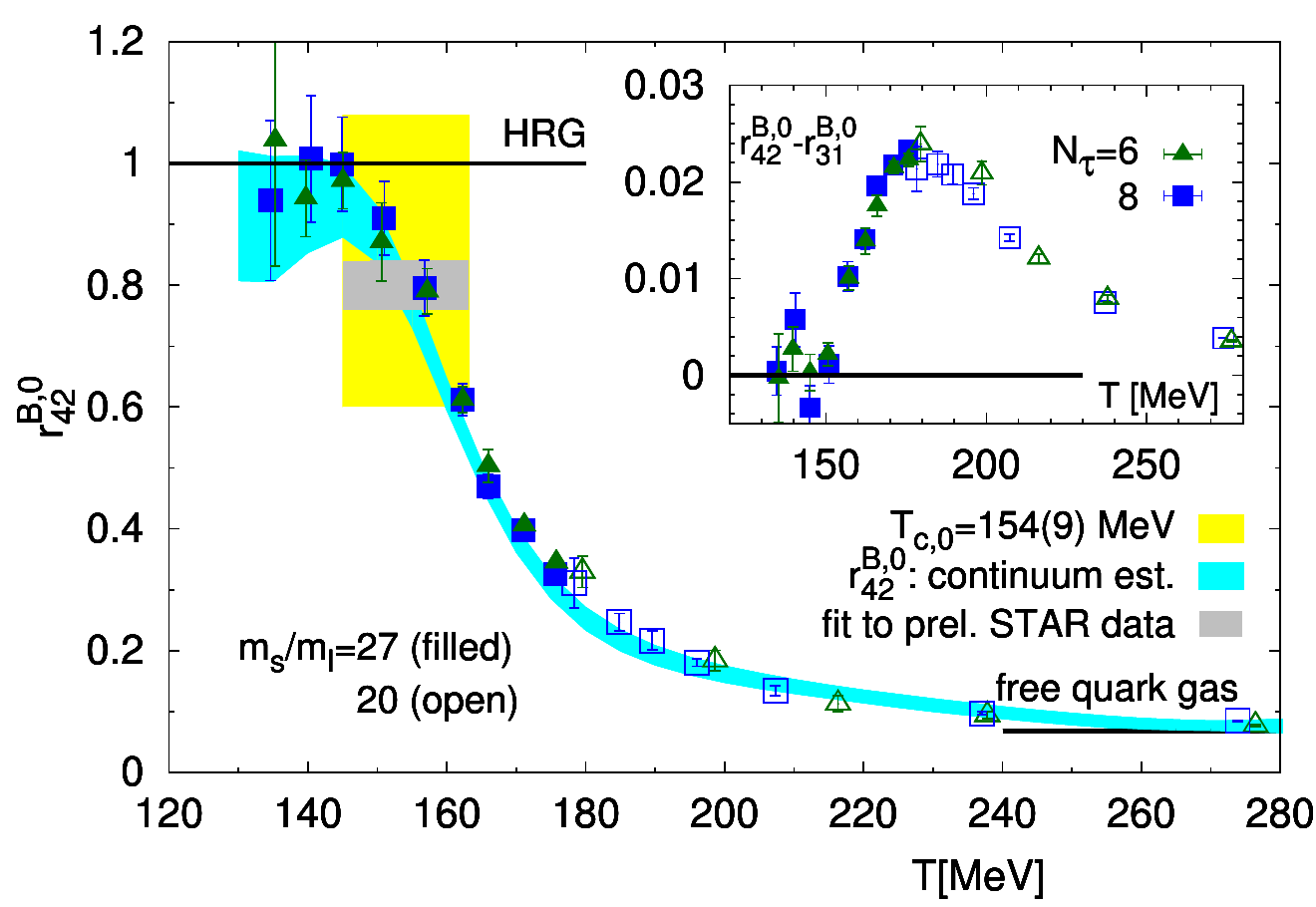}
\caption{Correlations between net baryon-number and net electric charge
as function of $\mu_B/T$ for three values of the temperature. Results from
a QCD calculation at next-to-leading order Taylor expansion are normalized
to a HRG model calculations truncated at the same order.}
\label{fig:BQmu}
\end{figure}

A consequence of the early deviation of $4^{th}$ order cumulants from
corresponding HRG calculations also is that differences between QCD
and HRG calculations increase with increasing value of the chemical 
potential and will show up already in $2^{nd}$ order cumulants. This is 
shown in Fig.~\ref{fig:BQmu}~(left) for the correlation
between net baryon-number and net electric charge evaluated up to
${\cal O}(\mu_B^2)$, 
\begin{equation}
\chi_{11}^{BQ}(T,\mu_B) = \chi_{11}^{BQ} + \frac{1}{2} \chi_{31}^{BQ} \mu_B^2
\; .
\label{BQ11}
\end{equation} 
Obviously 
QCD and HRG calculations differ significantly at $\mu_B/T=2$
already for $T\gsim 155$~MeV.

\section{Taylor expansions for skewness and kurtosis ratios} 

In HRG models with point-like non-interacting hadrons the distribution of 
net baryon 
number fluctuations is given by a skellam distribution. This leads to
quite simple properties of higher order cumulants. In particular,
the skewness ratio $S_B\sigma_B^3/M_B$ and the kurtosis ratio
$\kappa_B\sigma_B^2$ should both be equal to unity at all values of $\mu_B$. 
However, as discussed in the
previous section this cannot be expected to hold in QCD 
at temperatures above $T_{pc}$ where cumulants start to deviate 
significantly from HRG model calculations. The distribution of net 
baryon number fluctuations thus is e.g. not a simple skellam distribution.
QCD calculations of the skewness and kurtosis ratios yield
\begin{eqnarray}
\frac{S_B\sigma_B^3}{M_B} &=& \frac{\chi_3^B(T,\mu_B)}{\chi_1^B(T,\mu_B)}
= \frac{\chi_4^B +s_1 \chi_{31}^{BS}+ q_1 \chi_{31}^{BQ}}{\chi_2^B +s_1 \chi_{11}^{BS}+ q_1 \chi_{11}^{BQ}} +{\cal O}(\mu_B^2) \equiv r_{31}^{B,0}
+r_{31}^{B,2}\hat{\mu}_B^2 +{\cal O}(\mu_B^4)  \label{skewness} \\
\kappa_B \sigma_B^2 &=& \frac{\chi_4^B(T,\mu_B)}{\chi_2^B(T,\mu_B)} 
= \frac{\chi_4^B}{\chi_2^B} +{\cal O}(\mu_B^2) \equiv r_{42}^{B,0} 
+r_{42}^{B,2}\hat{\mu}_B^2+{\cal O}(\mu_B^4) \label{kurtosis}  \; ,
\end{eqnarray}
with $s_1$ and $q_1$ denoting the leading order expansion coefficients of
$\mu_S$ and $\mu_Q$ in terms of $\mu_B$ that result from imposing the
strangeness neutrality constraint $M_S=0$ and a fixed electric charge 
to baryon number ratio, $M_Q/M_B=0.4$. Results for the leading order
expansion coefficients, defined in Eqs.~\ref{skewness} and \ref{kurtosis}, 
are shown in Fig.~\ref{fig:BQmu}~(right).
While at $T\simeq 150$~MeV the kurtosis
ratio at $\mu_B=0$ is still close to unity,
$\kappa_B\sigma_B^2\simeq 0.9$, it strongly deviates from unity for
$T\simeq 160$~MeV, giving $\kappa_B\sigma_B^2\simeq 0.6$.  
In view of these differences it also is
not obvious that the skewness and kurtosis ratios  
will still coincide in a QCD calculation. However,
as can be seen from Eqs.~\ref{skewness} and \ref{kurtosis} at $\mu_B=0$ the 
leading order results will always be identical, if 
$\mu_Q=\mu_S=0$, irrespective of the 
size of deviation from the skellam limit. For non-zero $\mu_Q$ and 
$\mu_S$ the skewness and kurtosis ratios will differ. Nonetheless, as can 
be seen from the insertion in Fig.~\ref{fig:BQmu}~(right) this difference 
is small for all temperatures of interest. QCD thus predicts that the
skewness and kurtosis ratios will approach each other in the limit
$\mu_B\rightarrow 0$. This, however, changes for $|\mu_B|\ne 0$.

\begin{figure}[t]
\includegraphics[width=7.5cm]{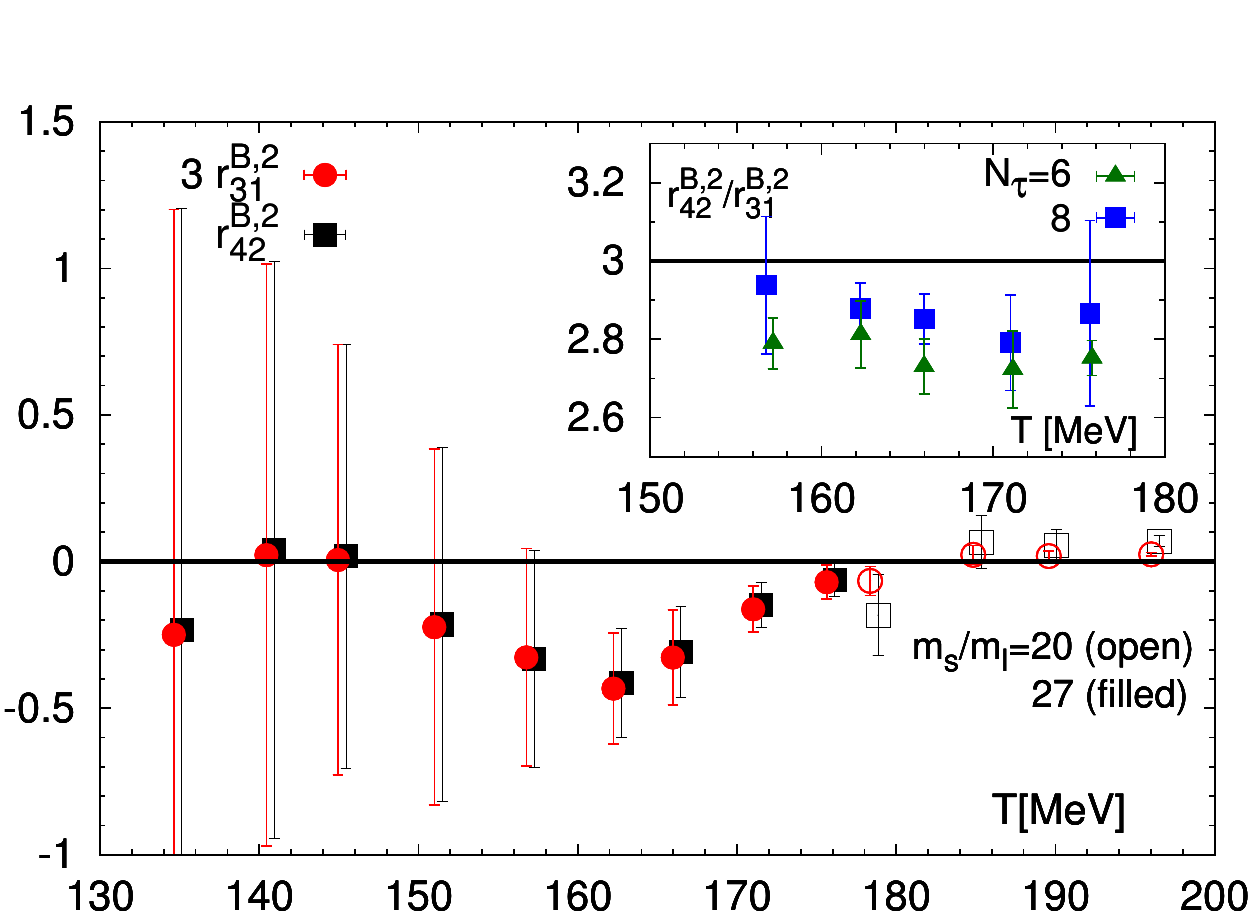}
\includegraphics[width=8.1cm]{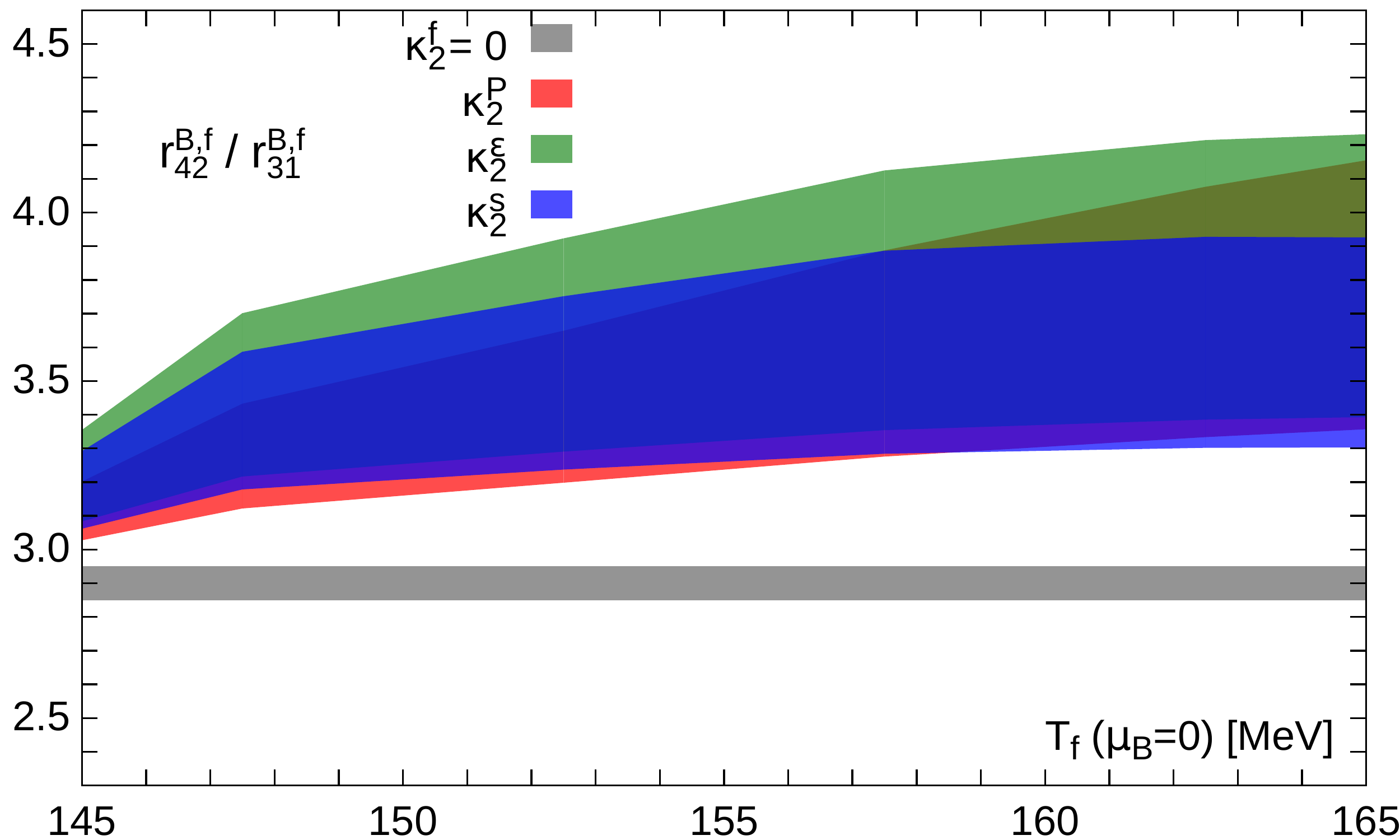}
\caption{{\it Left}:
The next-to-leading order (NLO) expansion coefficients of the skewness
and kurtosis ratios at fixed temperature as introduced in
Eqs.~\ref{skewness} and \ref{kurtosis}.
{\it Right}: The ratio of the NLO expansion coefficients taking into
account also a NLO expansion in terms of temperature along the line
of constant pressure, energy density or entropy density, respectively.
\label{fig:NLO}
}
\end{figure}

At next-to-leading order (NLO) the expansion coefficients $r_{31}^{B,2}$ and
$r_{42}^{B,2}$ need to be calculated. This is computationally difficult,
because $6^{th}$ order cumulants need to be evaluated. Current results for 
$r_{31}^{B,2}$ and $r_{42}^{B,2}$ 
are shown in 
Fig.~\ref{fig:NLO}~(left).
Although errors are still large, it is apparent that these expansion
coefficients are negative for $150~{\rm MeV} \lsim T\lsim 175~{\rm MeV}$
and that $r_{42}^{B,2}$ is about three times larger than $r_{31}^{B,2}$.
This has been reported by us earlier \cite{Karsch:2015nqx}. 
It thus is expected that the 
skewness and kurtosis ratios, which are almost identical at $\mu_B=0$,
will start to differ 
for $\mu_B>0$. As the expansion 
coefficients are negative, $\kappa_B\sigma_B^2$ will drop faster than
$S_B\sigma_B^3/M_B$. 

An additional subtlety in the analysis of the $\mu_B$-dependence of the skewness
and kurtosis ratios is that these need to be evaluated at a (freeze-out)
temperature that changes with $\mu_B$. This requires an additional Taylor
expansion of the ratios introduced in Eqs.~\ref{skewness} and \ref{kurtosis}.
Assuming that freeze-out happens along a line of constant physics
described either by constant pressure, energy density or entropy 
density, i.e. the lines shown in Fig.~\ref{fig:energy}, these contributions
can be evaluated using the Taylor expansion of the pressure \cite{inprep},
changing $r_{nm}^{B,2}$ to $r_{nm}^{B,f}$.
It turns out that the additional contributions increase the ratio
$r_{42}^{B,f}/r_{31}^{B,f}$ somewhat as shown in Fig.~\ref{fig:NLO}~(right).

\section{The STAR data on net proton-number skewness and kurtosis ratios}

\begin{figure}[t]
\includegraphics[width=7.5cm]{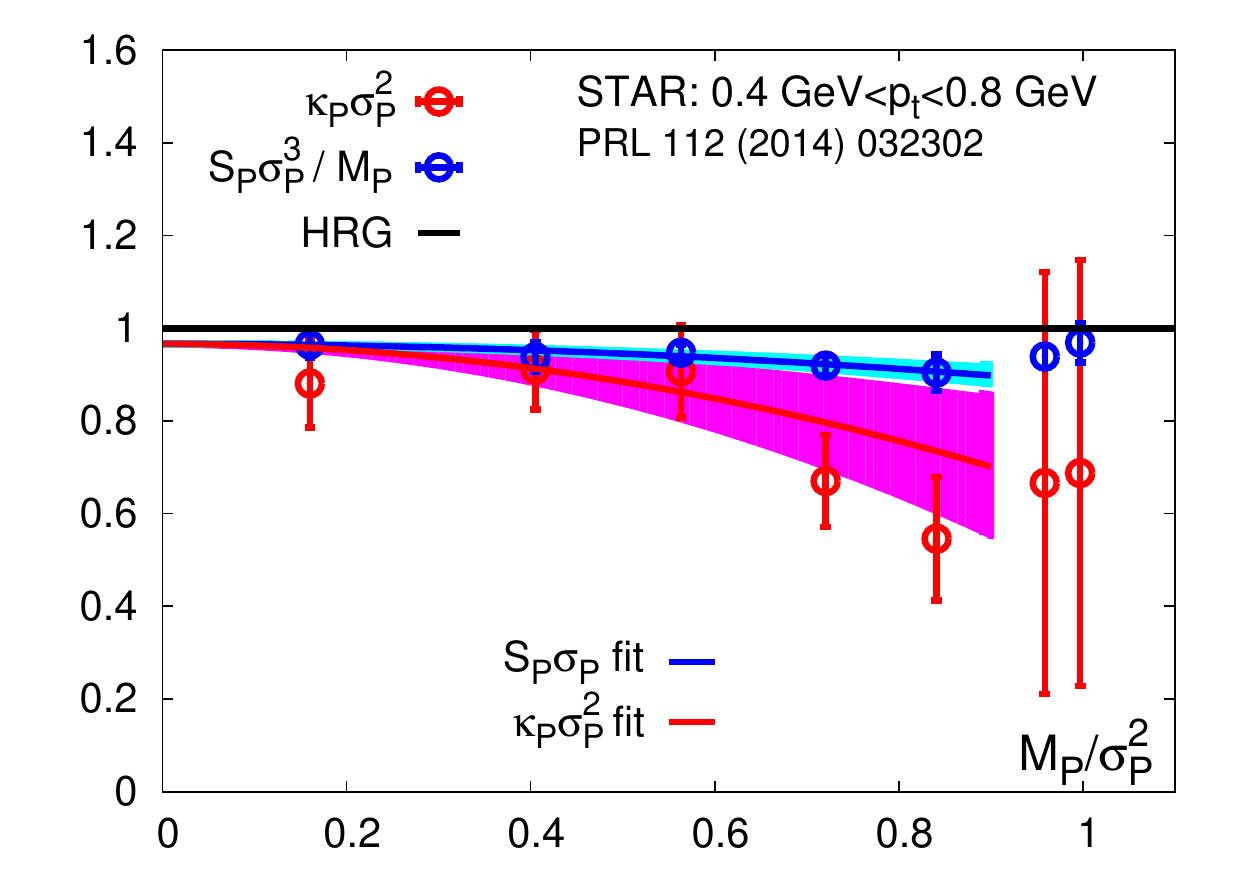}
\includegraphics[width=7.5cm]{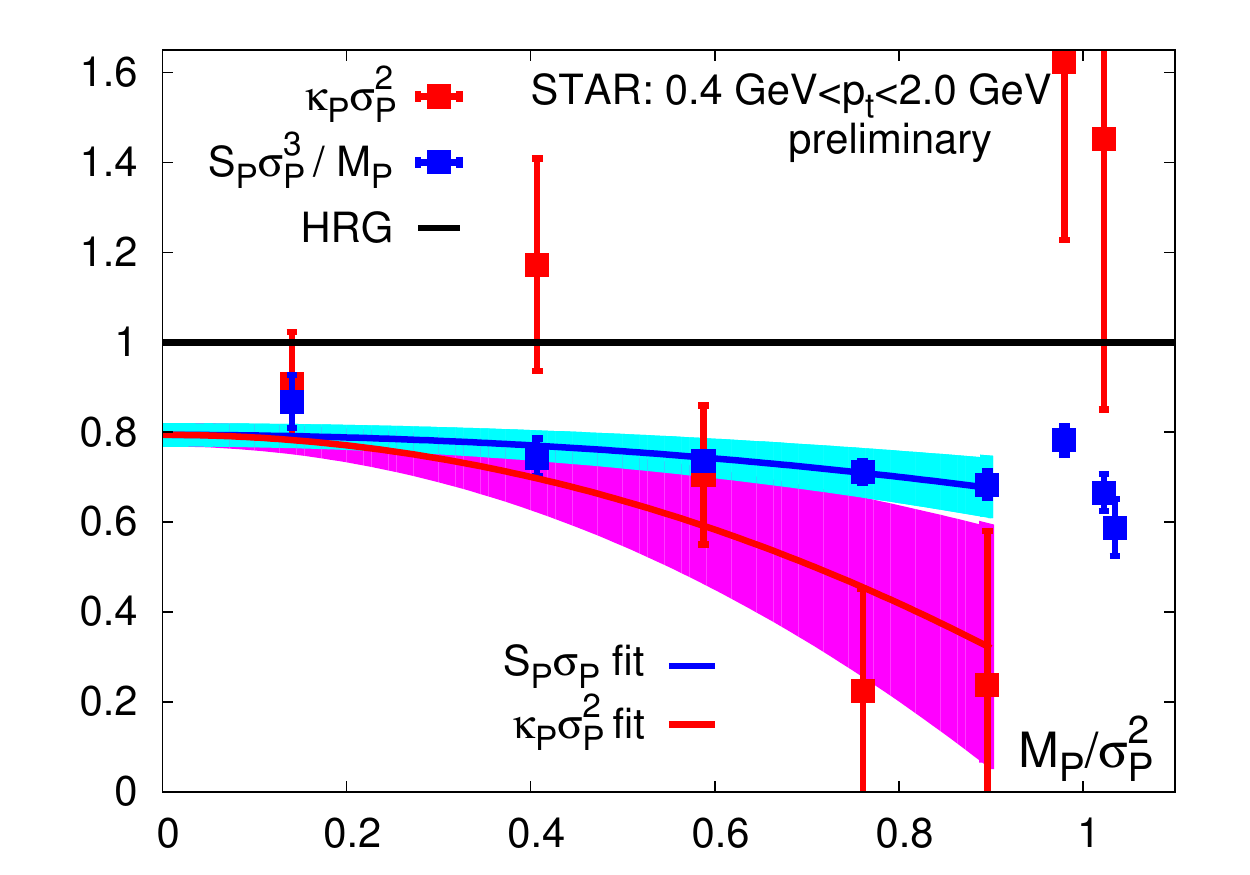}
\caption{Skewness and kurtosis ratios of net proton-number fluctuations
measured by the STAR collaboration in transverse momentum
intervals $0.4~{\rm GeV} < p_t > 0.8~{\rm GeV}$ (left) and
$0.4~{\rm GeV} < p_t > 2.0~{\rm GeV}$ (right), respectively. Data
are plotted versus the ratio of mean ($M_P$) over variance ($\sigma_P^2$) of 
the net proton-number distribution, which also is measured at various beam
energies. Curves show combined fits to the data for $M_P/\sigma_P^2 <0.9$
or equivalently $\sqrt{s_{NN}}\le 19.6$~GeV/fm$^3$. They are constraint by 
demanding $S_P\sigma^3/M_P=\kappa_P \sigma_P^2$ at $M_P/\sigma_P^2 =0$.
\label{data}
}
\end{figure}

The STAR collaboration has measured the skewness ratio $S_P\sigma_P^3/M_P$
and kurtosis ratio $\kappa_P \sigma_P^2$ of net proton-number fluctuations 
\cite{STARp08,STARp20}.
Obviously these cumulant ratios cannot directly be compared to QCD results on 
net baryon-number fluctuations. Already the strong sensitivity 
on the transverse momentum range used in the analysis, which is 
clearly visible in the data published so far by the STAR Collaboration
(Fig.~\ref{data}~(left) and (right), respectively), emphasizes that
there is need for better understanding of various effects that enter
the experimental analysis of higher order cumulants. Nonetheless, it is
striking that the published as well as the new preliminary data on 
skewness and kurtosis ratios resemble  all the features we expect to 
show up in equilibrium thermodynamics of QCD.
The ratios (i) are smaller than unity, (ii) they seem to coincide in 
the limit $\mu_B\rightarrow 0$, (iii) they have a negative slope
with increasing $\mu_B$ and (iv) the kurtosis ratio drops faster than the
skewness ratio. In fact, a combined quadratic fit to these ratios,
performed for all data obtained at beam energies 
$\sqrt{s_{NN}}\ge 19.6$~GeV/fm$^3$ and imposing the constraint 
$r_{42}^{B,0}/r_{31}^{B,0}$, yields for 
the ratio of slope parameters $r_{42}^{B,f}/r_{31}^{B,f} \sim 4\pm 2$, which
is in good agreement with the NLO QCD result shown in Fig.~\ref{fig:NLO}~(left).
These fits are shown in Fig.~\ref{data}.

\ack
This work has been partially supported through
the U.S. Department of Energy under Contract No. DE-SC0012704
and within the framework of the Beam Energy Scan Theory (BEST) 
Topical Collaboration,
and the German Bundes\-ministerium f\"ur Bildung und Forschung 
(BMBF) under grant no. 05P15PBCAA.

\end{document}